\documentstyle[psfig,emulateapj]{article}

\input psfig.sty

\lefthead{Goldberg \& Bacon}
\righthead{Second Order Lensing}

\begin{document}

\title{Galaxy-Galaxy Flexion: Weak Lensing to Second Order}

\author{David M. Goldberg \& David J. Bacon\\
Department of Physics, Drexel University, Philadelphia, PA 19104 ;
Institute for Astronomy, School of Physics, University of Edinburgh,
Royal Observatory, Edinburgh, EH9 3HJ, U.K.}
\begin{abstract}
In this paper, we develop a new gravitational lensing inversion
technique.  While traditional approaches assume that the lensing field
varies little across a galaxy image, we note that this variation in
the field can give rise to a ``Flexion'' or bending of a galaxy image,
which may then be used to detect a lensing signal with increased
signal to noise.  Since the significance of the Flexion signal
increases on small scales, this is ideally suited to galaxy-galaxy
lensing.  We develop an inversion technique based on the ``Shapelets''
formalism of Refregier (2003).  We then demonstrate the proof of this
concept by measuring a Flexion signal in the Deep Lens Survey.
Assuming an intrinsically isothermal distribution, we find from 
the Flexion signal alone a velocity width of $v_c=221\pm 12 km/s$
for lens galaxies of $r < 21.5$, subject to uncertainties in the
intrinsic Flexion distribution.

\end{abstract}


\section{Introduction}

The past several years have seen an explosion in the analysis of
weakly gravitationally lensed images of galaxies by galaxies
(e.g. Brainerd, Blandford \& Smail 1996; Hoekstra, Yee, \& Gladders, 2004),
clusters (e.g. Smail et al. 1997; Wittman, 2001; Gray et al. 2002;
Taylor et al. 2004), and Large Scale Structure (e.g. Wittman et
al. 2002; Hoekstra et al. 2002; Jarvis et al. 2003; Brown et al. 2003;
Pen et al. 2003; Bacon et al. 2003) producing an unprecedented glimpse
into the underlying matter distribution of the universe.
Galaxies-galaxy lensing, in particular, has proven a fertile testbed
of our understanding of structure formation.

Dynamical estimates of galaxy masses are subject to uncertainties
about whether a system is relaxed, and are limited by the lack of
luminous dynamical probes beyond a few tens of kpc from the center of
a galaxy.  Gravitational lensing techniques, on the other hand,
provide an accurate way of computing the surface density of distant
objects without recourse to dynamical estimates, since the distortion
of images is dependent only upon the potential field of a lens and not
upon either its composition or dynamics.  The success of these
endeavors has been such that there are a number of ongoing surveys of
weak lensing fields (e.g. Wittman et al. 2002, hereafter DLS; CFHT
Legacy Survey, {\tt http://www.cfht.hawaii.edu/Science/CFHTLS-DATA/})
and instruments (e.g. Advanced Camera for Survey, Clampin et al. 2000;
Dark Matter Telescope; Supernova/Acceleration Probe, Rhodes et
al. 2004) largely designed around the acquisition of large
high-quality lensing data.  Moreover, there remains a potential bounty
of information to be found in existing and ongoing survey data
(e.g. SDSS, York et al. 2000; Great Observatories Origins Deep Survey, 
Giavalisco et al. 2004; Medium Deep Survey, Ratnatunga, Griffiths \&
Ostrander 1999). Given the difficulty and expense in collecting
high-quality surveys of lensed galaxies it is imperative that we
extract as much information as possible from them.

One of the great advantages of studying weak lensing fields is that
the physics underlying gravitational lensing is well understood (see
e.g. Blandford \& Narayan, 1992; Kaiser \& Squires, 1993; Kaiser,
Squires \& Broadhurst, 1995, hereafter KSB; Mellier 1999; Bartelmann
\& Schneider, 2001 for reviews of weak lensing).  The KSB approach
provides the standard for most current analyses of weak lensing, and is
based upon estimating the ellipticity of a lensed galaxy as a probe of
the local shear field.  By measuring the ensemble properties of
ellipticity and orientation for a number of sources, one can make a
determination of the properties of the lens.

While the currently applied approaches have done an excellent job in
estimating the matter distribution of gravitational lenses, they
ultimately measure only the ellipticity of an image (the second
moments), and thus potentially drop significant information from
sources with significant substructure.  In order to improve on this,
we plan to extend the work of Goldberg \& Natarajan (2002) who
suggested that second order effects in gravitational lensing fields
may give rise to an octopole moment in the light distribution, which
expresses itself as ``flexion'' in the image.

The method of Goldberg \& Natarajan ultimately relied on a very
complicated form of the 2nd order shear operator, which made a
practical inversion difficult.  Instead, here we will cast this
approach into a ``Shapelets'' formalism (Refregier, 2003; Refregier \&
Bacon, 2003), a novel approach to both image and lensing analysis.
Rather than analyze image shapes according to their multipole moments,
shapelets methodology decomposes images into combinations of Hermite
polynomials.  As a reminder to the reader, the reduced Hermite
polynomials are the eigenfunctions of the simple harmonic oscillator
in quantum mechanics, $B_n(x)$.  These functions have a number of
useful properties, including orthogonality and a Gaussian factor,
which localizes the function.

In \S~\ref{sec:lensing}, we begin by reviewing basic properties of the
``forward'' problem in weak gravitational lensing, and define our
notation within the present work.  We then proceed to introduce the
second order term in the lensing operator, and then discuss a
particularly useful approach to the second order problem using the
shapelets formalism.  In \S~\ref{sec:signal} we discuss a practical
technique for inversion of the density gradient signal. In
\S~\ref{sec:DLS} we provide a proof of concept by measuring the second
order shear signal from Galaxy-Galaxy lensing in several DLS shear
fields.  We conclude with a discussion of future prospects.

\section{Second Order Lensing}
\label{sec:lensing}

\subsection{Review of Weak Lensing Formalism}

We begin with a brief review of weak lensing theory.  An excellent
discussion of this material can be found in Bartelmann \& Schneider
(2001), from which we borrow our conventions.  Imagine that we are
observing a set of extended sources at an angular diameter distance,
$D_s$ which are gravitationally lensed by a mass distribution at an
angular diameter distance, $D_l$.  We thus define a dimensionless
surface density, the convergence, $\kappa$, such that:
\begin{equation}
\kappa({{\bf x}})\equiv \frac{D_{ls}D_l}{D_s} \frac{4\pi G
\Sigma({\vec\theta})}{c^2}\ ,
\label{eq:kappa_def}
\end{equation}
where $\Sigma$ is the surface density of the lens, $D_{ls}$ is the
distance between lens and source and ${\bf x}$ represents the image
coordinates as seen by the observer (neglecting a constant coordinate
transform).

The convergence may be thought of as a source term for a potential,
$\psi({\bf \theta})$, and related via a Poisson-like equation:
\begin{equation}
\nabla^2 \psi({\bf x})=2\kappa({\bf x})\ ,
\label{eq:psi_def}
\end{equation}
where all gradients and divergences are calculated in two dimensions.

Since lensing conserves surface brightness, a mapping from foreground
to background coordinates is sufficient to determine a background
brightness map from a foreground one (or vice-versa) provided a full
knowledge of the geometry of the system (cosmology plus the redshifts
of the source and lens) and mass distribution of the lens. Thus, we
may expand around the origin to determine a deprojection operator on a
foreground light distribution, which yields the amplification matrix,
\begin{equation}
{\bf A}({{\bf x}})\equiv
\frac{\partial{{\bf x}'}}{\partial {{\bf x}}}=
\left( 
\delta_{ij}-\frac{\partial^2\psi({{\bf x}})}{\partial
x_i\partial x_j}\right)\equiv
\left(
\begin{array}{cc}
1-\kappa-\gamma_1 & -\gamma_2 \\
-\gamma_2 & 1-\kappa+\gamma_1
\end{array}
\right)\ .
\label{eq:Adef}
\end{equation}
where ${\bf x}'$ are the coordinates that the lensed image would have
in the absence of lensing.  Note that for convenience we set the
origins of both the foreground and background coordinate systems to be
the centers of light in their respective planes.  

Equation~(\ref{eq:Adef}) is the first term in a Taylor series expansion
of the distortion operator. The term $\gamma$ is a complex shear term,
representing the anisotropic part of the distortion, with
$\gamma=|\gamma|e^{2i\phi}$, and the real and imaginary parts being
denoted with the subscripts, ``1'' and ``2'' respectively, as per
convention.  

If we assume that the convergence and shear field were constant along
the scale of the lensed image, the brightness map of a galaxy would
transform with the simple relation:
\begin{equation}
x'_i=A_{ij}x_j\ .
\end{equation}

\subsection{Expansion to 2nd Order}

Of course, the field is not a constant.  Thus, we may imagine
expanding the field to second order such that:
\begin{equation}
x'_i\simeq A_{ij}x_j +\frac{1}{2}D_{ijk} x_j x_k
\end{equation}
where 
\begin{equation}
D_{ijk}=\frac{\partial A_{ij}}{\partial x_k}
\end{equation}
As Kaiser (1995) has pointed out, there are a number of relationships in the
derivatives of $\kappa$ and $\gamma$ which greatly simplify the
construction of ${\bf D}$:
\begin{equation}
\left( \begin{array}{c} 
\kappa_{,1} \\
\kappa_{,2}
\end{array}
\right)=
\left( \begin{array}{c} 
\gamma_{1,1}+\gamma_{2,2} \\
\gamma_{2,1}-\gamma_{1,2}
\end{array}
\right)\ .
\label{eq:kaiser}
\end{equation}
Thus,
\begin{eqnarray}
D_{ij1}&=&\left(
\begin{array}{cc}
-2\gamma_{1,1}-\gamma_{2,2} & \ \ -\gamma_{2,1} \\
-\gamma_{2,1} & \ \ -\gamma_{2,2} 
\end{array}
\right) \\ \nonumber
D_{ij2}&=& \left(
\begin{array}{cc}
-\gamma_{2,1} & \ \ -\gamma_{2,2}\\
-\gamma_{2,2} & \ \ 2\gamma_{1,2}-\gamma_{2,1}
\end{array}
\right)
\label{eq:ddef}
\end{eqnarray}

Now, let us suppose that there exists some background brightness
field, $f({\bf x}')$, such as the brightness of some background
galaxy.  Since lensing preserves surface brightness, we may write the
apparent foreground brightness field as:
\begin{equation}
f({\bf x})=f'[{\bf x}'({\bf x})]\ .
\end{equation}
We can write the brightness as:
\begin{eqnarray}
f({\bf x})&\simeq&f'({\bf A}{\bf x} +
	\frac{1}{2}{\bf D} {\bf x} \otimes {\bf x}) \\ \nonumber
&=&f'[{\bf x}+({\bf A}-\hat{\bf I}){\bf x}+
	\frac{1}{2}{\bf D} {\bf x} \otimes {\bf x}] \\ \nonumber
\end{eqnarray}
Or, expanding as a Taylor series:
\begin{eqnarray}
f({\bf x})&\simeq&\left\{1+
	\left[ (A-I)_{ij}x_j+\frac{1}{2}D_{ijk}x_jx_k\right]
	\frac{\partial}{\partial x_i} \right.\\ \nonumber
	&+&\left. \frac{1}{2}
	\left[ (A-I)_{ik}x_k+\frac{1}{2}D_{ikl}x_k x_l\right]
	\left[ (A-I)_{jk}x_k+\frac{1}{2}D_{jkl}x_k x_l\right]
	\frac{\partial^2}{\partial x_i \partial x_j} \right\}
	f'({\bf x}) \\ \nonumber
\label{eq:ftaylor}
\end{eqnarray}
Expanding out the terms on the first line the expressions are first
order in $\gamma$, and (up to) 2nd order in position, while the terms
on the second line are (at least) 2nd order in position, and 2nd order
in $\gamma$.  If the scale on which $\gamma$ varies is typically
smaller than the image size, the terms on the second line will
necessarily be smaller than the 2nd order contribution on the first,
and thus we will consider only the first line.  

Thus, in the limit of a smoothly varying, weak shear field, we have
the relation:
\begin{equation}
f({\bf x})\simeq\left\{1+
	\left[ (A-I)_{ij}x_j+\frac{1}{2}D_{ijk}x_jx_k\right]
	\frac{\partial}{\partial x_i} \right\}	f'({\bf x}) \ .
\label{eq:2ndorder}
\end{equation}

\subsection{Image Analysis Using Shapelets}

While equation~(\ref{eq:2ndorder}) defines a linear transformation on
an image due to lensing, in practice it is fairly complicated to
actually perform such a transformation (or the inversion) in
generality.  However, we may simplify this problem considerably by
decomposing the image into basis coefficients.

Following Refregier \& Bacon (2003), we expand the image into Reduced
Hermite polynomials, or ``Shapelets.''  The light distribution,
$I(x,y)$ of a galaxy is expanded as a combination of two-dimensional
basis functions:
\begin{equation}
f({\bf x})=\sum_{n,m} f_{nm} B_{nm}({\bf x}) \ .
\end{equation}
where
\begin{equation}
B_{nm}({\bf x};\beta)=\beta^{-1}\phi_n(\beta^{-1}x_1)\phi_m(\beta^{-1}x_2)\ ,
\end{equation}
and where $\beta$ is a scaling factor, and $\phi_n$ are the reduced
Hermite polynomials:
\begin{equation}
\phi_n(x)=\left[2^n\pi^{1/2}n!\right]^{-1/2}{\cal H}_n(x)e^{-\frac{x^2}{2}}
\end{equation}
such that:
\begin{equation}
{\cal H}_n''-2x{\cal H}_n'+2n{\cal H}_n=0\ .
\end{equation}

Since the lowest order polynomials resemble Gaussian light profiles,
Refregier (2003) demonstrates that for typical HST images, convergence
can rapidly be reached using a few$\times 10$ coefficients.

It should be noted that though this expansion is not the same as a
multipole expansion, it has many similar properties.  For example, an
image with $f_{20}=a^2$, and $f_{02}=b^2 < a^2$ will look like an
ellipse with a Gaussian radial profile.  Likewise, any combinations
for which $n+m$ is even produces an image which is symmetric with
respect to 180 degree rotations (see, e.g. Figure 2 in Refregier,
2003).  The canonical pictures of both spiral and elliptical galaxies
have precisely this symmetry, and thus, it is possible that typical
galaxies can be reconstructed almost exclusively from even moments.

If, taking our example from the SHO in quantum mechanics, we define:
\begin{equation}
\hat{p}_i\equiv \frac{\partial}{\partial x_i} \ \ ; \ \ \hat{x}_i=x_i
\end{equation}
and expand Equation~\ref{eq:2ndorder} to first order in $\gamma$, then
we get:
\begin{equation}
f({\bf x})\simeq \left[1+(A-I)_{ij}\hat{x}_j \hat{p}_i+
\frac{1}{2}D_{ijk}\hat{x}_j\hat{x}_k \hat{p}_i \right]
f'({\bf x})
\label{eq:2ndoperator}
\end{equation}

Again following Refregier \& Bacon (2003), we expand our light
function in Shapelets and can apply the distortion operator as a
combination of raising and lowering operators.  Note that we have
implicitly defined our coordinates such that the $\beta$ parameter
used in Refregier (2003) is equal to 1.  Recall,
\begin{eqnarray}
\hat{x}_i=\frac{1}{\sqrt{2}}(\hat{a}_i+\hat{a}_i^\dagger) & \ \ \ & 
\hat{p}_i=\frac{1}{\sqrt{2}}(\hat{a}_i-\hat{a}_i^\dagger)
\end{eqnarray}
and can operate on a Shapelet (e.g. with $i=1$) as:
\begin{equation}
\hat{a}_1 \left| \phi_{n\ m}\right>=\sqrt{n} \left| \phi_{n-1\  m} \right>
\end{equation}
\begin{equation}
\hat{a}^\dagger_1 \left| \phi_{n\ m}\right>=\sqrt{n+1} \left|
\phi_{n+1\ m} \right>
\end{equation}
and similarly for the y-axis.

Now, the first thing to note is that with the addition of the $\hat D$
operator, there are odd combinations of $\hat x$ and $\hat p$, meaning
that there can be coupling between $\Delta n+\Delta m=$ odd modes.
Contrary to ordinary weak lensing, in which an image is lensed
symmetrically around its center of light (and thus, the centroid
remains fixed), second order effects will cause a shift in the center
of light.  We must thus compute the shift in the center of light.

Because of the inherent symmetries in the linear lensing operator,
the only contribution to the centroid shift comes from the second
order operator:
\begin{equation}
\langle x_l \rangle=\frac{1}{2} D_{ijk} \int d^2{\bf x}\  x_l x_k x_k
\frac{df'}{dx_i}
\label{eq:pos_shift}
\end{equation}
Expanding this out and integrating, we find that the center of light
in the foreground will be shifted by:
\begin{eqnarray}
\langle x_1 \rangle &=&-\langle x_1^{'2} \rangle \left(
\frac{3}{2} D_{111}+D_{212}\right)-\langle x'_1 x'_2 \rangle
(2D_{112}+D_{222})-\langle x_2^{'2}\rangle
\left(\frac{1}{2}D_{122}\right) \\ \nonumber
\langle x_2 \rangle &=& -\langle x_1^{'2}\rangle
\left(\frac{1}{2} D_{211}\right)-\langle x_1' x_2'\rangle \left(
2 D_{212}+D_{111}\right)-\langle x_2^{'2}\rangle
  \left(\frac{3}{2}D_{222}+D_{112}\right)
\end{eqnarray}
where the second moments of the light distribution can be computed
from the foreground field (since changes in the moments from
background to foreground will necessarily be 2nd order in $\gamma$,
and thus negligible).  

A translational shift can be given by the operator:
\begin{equation}
\hat{T}_i=\frac{1}{\sqrt{2}}(\hat{a}_i^\dagger -\hat{a}_i)
\label{eq:translational}
\end{equation}

Combining all of this yields the following lensing operator:
\begin{equation}
f({\bf x})\simeq (1+\kappa \hat{K}+\gamma_i
\hat{S}^{(1)}_i+\gamma_{i,j}\hat{S}^{(2)}_{ij}) f({\bf x}')
\label{eq:lens}
\end{equation}
where $\hat{K}$ is the convergence operator:
\begin{equation}
\hat{K}=1+\frac{1}{2}(\hat{a}_1^{\dagger 2}+\hat{a}_2^{\dagger
  2}-\hat{a}_1^2-\hat{a}_2^2)
\end{equation}
 and $\hat{S}^{(1)}_i$ is
the linear shear operator defined by Refregier (2003):
\begin{eqnarray}
\hat{S}^{(1)}_1&=&\frac{1}{2}(\hat{a}_1^{\dagger 2}-\hat{a}_2^{\dagger
  2}-\hat{a}_1^2+\hat{a}_2^2)\\
\hat{S}^{(1)}_2&=&\hat{a}_1^\dagger\hat{a}_2^{\dagger}-\hat{a}_1\hat{a}_2\ .
\end{eqnarray}
It is straightforward, albeit tedious, to derive the explicit form of
the second order operator, $\hat{S}^{(2)}$ from
equations~(\ref{eq:2ndoperator}) and
(\ref{eq:pos_shift}-\ref{eq:translational}).  We write out the
explicit 2nd order transformation as:
\begin{eqnarray}
S_{11}^{(2)}&=&\frac{1}{4\sqrt{2}}\left[-2
  \hat{a}_1^3+\hat{a}_1\left(4-2\hat{N}+12<xx>\right)+
8<xy>\hat{a}_2-8<xy>\hat{a}_2^{\dagger}+
\hat{a}_1^{\dagger}\left(6+2\hat{N}-12<xx>\right)\right.\\
  \nonumber
& & \left. +2 \hat{a}_1^{\dagger 3}\right] \\
S_{12}^{(2)}&=&\frac{1}{4\sqrt{2}}\left[-8<xy>
  \hat{a}_1+2\hat{a}_2^3+
\hat{a}_2\left(-4+2\hat{M}-12<yy>\right)+
\hat{a}_2^{\dagger}\left(-6-2\hat{M}+12<yy>\right)\right.\\ \nonumber
&&\left.-2\hat{a}_2^{\dagger 3}+8<xy>\hat{a}_1^{\dagger}\right]\\
S_{21}^{(2)}&=& \frac{1}{4\sqrt{2}}\left[-3 \hat{a}_1^2\hat{a}_2
-\hat{a}_1^2\hat{a}_2^{\dagger}
+12<xy>\hat{a}_1
-\hat{a}_2^3
+\hat{a}_2\left(3-2\hat{N}-1\hat{M}+2<xx>+10<yy>\right)\right.\\ \nonumber
&&\left.+\hat{a}_2^{\dagger
  1}\left(6+2\hat{N}+\hat{M}-2<xx>-10<yy>\right)
+\hat{a}_2^{\dagger 3}
-12<xy>\hat{a}_1^{\dagger}
+\hat{a}_1^{\dagger 2}\hat{a}_2+3\hat{a}_1^{\dagger 2}\hat{a}_2^{\dagger}\right]\\
S_{22}^{(2)}&=&\frac{1}{4\sqrt{2}}\left[-\hat{a}_1^3
-3\hat{a}_1\hat{a}_2^2
+\hat{a}_1\left(3-\hat{N}-2\hat{M}+10<xx>+2<yy>\right)
+\hat{a}_1\hat{a}_2^{\dagger 2}+12<xy>\hat{a}_2 \right.\\ \nonumber
&&\left.-12<xy>\hat{a}_2^{\dagger}
 -\hat{a}_1^{\dagger}\hat{a}_2^2
+\hat{a}_1^{\dagger}\left(6+\hat{N}+2\hat{M}-10<xx>-2<yy>\right)
+3\hat{a}_1^{\dagger}\hat{a}_2^{\dagger 2}+\hat{a}_1^{\dagger
  3}\right] \ ,
\end{eqnarray}
where $\hat{N}B_{nm}(x,y)=nB_{nm}(x,y)$, and similarly for $\hat{M}$.

Interestingly, all of the terms in the second order shear couples
$\Delta n+\Delta m=$odd coefficients, while the first term shear and
convergence operators couple $\Delta n+\Delta m = 2$.  In other words,
we may consider only the ``odd'' (meaning n+m=odd) moments when
estimating the parameters $\gamma_{i,j}$, and only the even moments
($n+m=$even) when estimating the shear $\gamma_i$, since all
coefficients with $n$ {\it or} $m$ as odd in the background image will
have an expectation value of zero.  Since deviations from zero suggest
a second order signal, we are not required to subtract two large modes
(e.g. $f_{20}-f_{02}$), and thus add considerable noise, in order to
extract a signal.  We discuss a practical inversion technique in the
next section.

\section{Practical Parameter Estimation}
\label{sec:signal}

While the above relations appear fairly complicated, in reality,
inversion is actually quite straightforward.  Below, we describe a
practical technique of inverting a set of shapelet coefficients to
yield an estimate of the shear and its derivatives.

\subsection{Measurement and Signal Noise}

As Refregier (2003) notes, if the pixel noise is independent
(e.g. Poisson), then the covariance matrix of shapelet coefficients
will be be simply related by:
\begin{equation}
V_{{\bf n}_1 {\bf n}_2}=\sigma_N^2\delta_{{\bf n}_1 {\bf n}_2}\ ,
\end{equation}
where $\sigma_N$ is simply the Poisson noise of the integrated signal.
Though each mode has equal noise, the signal strength from the first
and second order lensing signals are quite different, and it would be
helpful to consider in which regime each dominates.  

The approximate strength of the second order lensing signal may be
inferred from equation~(\ref{eq:2ndorder}).  Note that all of the
terms in $D_{ijk}$ are proportional to $d\gamma/dr$.  If the lens is
an isothermal sphere (or any other power law relation), then
$D_{ijk}\propto \gamma/R_{lens}$, where $R_{lens}$ is the angular
distance from the source to lens.  Likewise, the integral of $x_j x_k
\partial/\partial x_i$ over the source image will produce a term
proportional to the angular scale of the image, itself.  We give the
``size'' of a galaxy image as its semi-major axis, $a_{gal}$.  Thus,
the 2nd order lensing signal will have a strength which scales like:
\begin{equation}
\delta f_{nm=odd}\propto \gamma \frac{a_{gal}}{R_{lens}}\ , 
\end{equation}
which means that unless the source is relatively close to the lens in
the image plane, measurement errors can swamp the signal.  However, it
is anticipated (and below, shown), that the {\it intrinsic} variance
in this signal is significantly smaller than the variance in the
shear.  Moreover, in the particular regime of interest -- nearby
galaxy pairs, as found in galaxy-galaxy fields, this signal will be
particularly well-pronounced.

It should also be noted that, in general, the observed galaxy image is
convolved with a PSF.  Refregier \& Bacon (2003) discuss a PSF
inversion technique.  We do not apply the inversion in the present
work, restricting ourselves to large galaxies where the impact of the
PSF is small.  However, we will explicitly perform a PSF inversion in a
forthcoming paper.  It should further be noted that inversion of the
PSF gives rise to a non-diagonal covariance matrix, and we thus use
the more general form of the covariance matrix in our parameter
estimation below.

\subsection{Inversion of the Shapelet Coefficients}

In order to compute the lensing coefficients, we need to invert
equation~(\ref{eq:lens}).  In practice, this is somewhat simpler than
it might initially appear, since, for example, the convergence,
$\kappa$ cannot be uniquely estimated from a given shear field (a
constant $\kappa$ value can be arbitrarily added to the reconstruction).
This is known as the ``mass-sheet degeneracy,'' (Kaiser \& Squires
1993), and, excluding this effect, we may write down a goodness of fit
relation as:
\begin{equation}
\chi^2\equiv\left[\mu_{n_1m_1}-f_{n_1m_1}+(\gamma_i \hat{S}_i^{(1)}+\gamma_{i,j}S_{ij}^{(2)})
  \overline{f}_{n'_1m'_1}\right]V^{-1}_{n_1m_1 n_2
  m_2}
 \left[(\mu_{n_2m_2}-f_{n_2m_2}+(\gamma_i \hat{S}_i^{(1)}+\gamma_{i,j}S_{ij}^{(2)})
  \overline{f}_{n'_2m'_2}\right]
\end{equation}
where the lensing operators implicitly bring power from the primed to
unprimed indices.  We have defined, $\mu_{nm}$ as the ``expected''
source signal.  For $n+m$=even, this is simply the average of the
measured signal (since the universe has no preferred direction), and
for $n+m$=odd, we set this to zero.  Likewise, $\overline{f}_{nm}$ is the
best estimate unlensed signal.  This is subtly different as, though
the odd terms are still set to zero, the even terms are the observed
coefficients.  Since lensing (especially in galaxy-galaxy fields) is
expected to be small, the best estimate of an intrinsic moment is the
measured moment itself.  

Because we have defined the source terms for the odd moments as zero,
we may break the above expression into two separate terms:
\begin{eqnarray}
\chi^{2(even)}&=&\left[\mu_{n_1m_1}+(\gamma_i \hat{S}_i^{(1)}-1)
  f_{n'_1m'_1}\right]V^{-1}_{n_1m_1 n_2
  m_2}
 \left[(\mu_{n_2m_2}-(1+\gamma_i \hat{S}_i^{(1)})
  f_{n'_2m'_2}\right]\\
\chi^{2(odd)}&=&\left[f_{n_1m_1}-\gamma_{i,j}S_{ij}^{(2)}
  f_{n'_1m'_1}\right]V^{-1}_{n_1m_1 n_2
  m_2}
 \left[(f_{n_2m_2}-\gamma_{i,j}S_{ij}^{(2)}f_{n'_2m'_2}\right]
\label{eq:inverse}
\end{eqnarray}
Since the first order term only lenses even moments to even moments,
and the second order term only lenses even moments to odd moments, the
two can be computed independently.  In other words, the second-order
effects represent an entirely new calculation, while the first order
$\chi^2$ can be minimized exactly as described by Refregier \& Bacon
(2003).

Once the gradient of the complex shear has been estimated via
minimization of $\chi^2$, we may relate this to the gradient of the
convergence using equation~(\ref{eq:kaiser}).  We refer to the bending
of the observed image as the estimated ``Flexion'':
\begin{equation}
{\bf {\cal F}}\equiv (\tilde{\gamma}_{1,1}+\tilde{\gamma}_{2,2}){\bf
  i}+
(\tilde{\gamma}_{2,1}-\tilde{\gamma}_{1,2}){\bf j}\ ,
\end{equation}
where $\tilde{\gamma}_{i,j}$ is the estimated inversion of the shear
derivatives from the $\chi^2$ minimization.  Note that, in principle,
${\bf {\cal F}}$ could be used to reconstruct a convergence field.  In
the present worked example, however, noise plays too much of a role. 

\subsection{Noise}

With no atmospheric or instrumental effects, the gradient estimator
can be related to the true signal via the relation:
\begin{equation}
{\bf {\cal F}} = \nabla \kappa \pm \sigma_{{\bf {\cal F}}, P} \pm
\sigma_{{\bf {\cal F}}, S}
\end{equation}
We have already discussed the Poisson measurement noise ($\sigma_{{\bf
{\cal F}}, P}$) above.  However, much like in conventional lensing, we
must also take into account the intrinsic scatter in shapes amongst
real galaxies.  

To measure the intrinsic scatter, we observed two complementary
regimes.  First, we measured the shear and flexion variance in the DLS
fields themselves. Secondly, we examined the shear and flexion in two
clusters from the HST archive, Abell 665 and Abell 2390, and selected
elliptical and spiral galaxies.  We found 75 ellipticals and 53
spirals in our claster sample which were large and bright enough to
classify by eye.  No classification was done on DLS galaxies.

\begin{figure}
\centerline{\psfig{figure=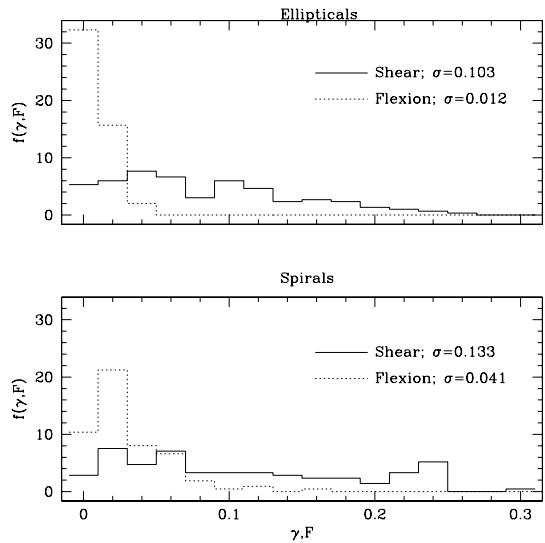,angle=0,height=3.0in}}
\caption{The fraction of observed galaxies with a particular measured
  ellipticity (solid) and flexion, $a_{gal}\times {\bf {\cal F}}$
  (dashed) from HST clusters A665, and A2390.  Since lensing effects
  are generally small, and lensing orientation angles are randomized,
  we suppose this distribution to represent the intrinsic galaxy shape
  distribution of the background sample.  Note that while the shear
  distribution for both spirals and ellipticals are similar,
  ellipticals produce a far smaller scatter in Flexion than do spirals.}
\label{fg:distribution}
\end{figure}

For these samples, we measure the distribution function of ellipticity
(shear) and Flexion; the results for clusters are shown in
Fig.~\ref{fg:distribution}.  We define the Flexion in units of the
inverse of the semi-major of the observed galaxy.  In this way, it is
a dimensionless, and distance-independent measure of the shape.
We find that for both ellipticals and spirals, the standard deviation
in shear is relatively low, $\sigma_{\gamma,E}=0.10$, and
$\sigma_{\gamma,S}=0.13$, respectively.

The scatter in the shear is somewhat lower than that which is normally
measured.  For example, in the COMBO-17 sample discussed in Brown et
al. (2003), the standard deviation of the ellipticity is approximately
0.25, approximately twice as large, and consistent with the estimate
from Brainerd et al. (1996).  However, taking the same sample, and
using only those galaxies with $r<17$, $\sigma_\gamma=0.16$.  Since
traditional approaches use the KSB technique to invert the PSF for
small galaxies (an intrinsically noisy inversion), additional scatter
is produced in the shear.  Since we are interested in the intrinsic
scatter for the present work, we use our lower estimate of
$\sigma_\gamma$.  We caution, however, that, even compared to bright
galaxy sample in COMBO-17, this may be a modest underestimate of the
true value.

For the Flexion, we find a much larger scatter among spirals than
ellipticals, with a standard deviation of $\sigma_{F,S}=0.041$, and
$\sigma_{F,E}=0.012$, respectively.  This gives the expected result
that early-type galaxies are significantly more regular than late
types.

Since sources with lower resolution will be harder to classify, we
would like to take a typical Flexion and shear for field galaxies.
Making the approximation that 60\% of a randomly selected background
population will be spirals (e.g. Postman \& Geller 1984), we find
a typical shear of $\sigma_\gamma=0.12$, and $\sigma_F=0.029$.  These
results have a similar to that found in the well resolved
(but morphologically unclassified) sample from the DLS, yielding,
$\sigma_{\gamma,DLS}=0.14$, and $\sigma_{F,DLS}=0.04$.  

Using the HST estimates, we may perform a simple estimate of signal to
noise.  For the first order signal, we find:
\begin{equation}
\left(\frac{S}{N}\right)^{(1)}\simeq \frac{\gamma}{0.12}
\end{equation}
And, likewise, for the flexion signal,
\begin{equation}
\left(\frac{S}{N}\right)^{(2)}\simeq \frac{\gamma / R_{lens}}{0.029 /
  a_{gal}}\ ,
\end{equation}
since for an isothermal sphere
\begin{eqnarray}
\frac{d\kappa}{dr}&=&-\frac{d\gamma}{dr}\\ \nonumber
&=&\frac{\gamma}{R_{lens}}
\end{eqnarray}
since $\gamma\propto 1/R$.

The signals from first and second order effects will be comparable if:
\begin{equation}
R_{lens} \simeq 4.1 a_{gal}
\label{eq:intrinsic}
\end{equation}
Thus, at modest separations, the signal to noise from the second order
signal will be comparable to the traditional shear signal.  This ratio
will rise to a factor of 10 if the background sample consists entirely
of ellipticals.  Moreover, it provides an entirely new source of
information, as it is a direct estimator of the underlying surface
density of the lens.

\subsection{Parameter Estimation with a Nearly Circular PSF}

In practice, we do not generally directly estimate either the shear or the
flexion directly from averages over the ellipticities.
Even a perfectly circular PSF will alter the value, but not the
orientation, of the estimated parameters.  As a result, we will
generally wish to consider only the relative orientation of the shear
or the flexion with respect to the candidate lens.  

In practice, this means measuring the Flexion and shear of a large
number of background sources which separated from their respective
lenses by a fixed range of distances.  The distribution function of
the relative orientation angles of both Flexion and shear and then
computed.  In the absence of lensing, we would assume these
distributions would be drawn from a uniform prior.  However, lensing
will tend to orient the shear perpendicular to the lens, and the
Flexion toward the lens.

We must thus relate the distribution function of relative orientation
angles, $P_\phi$, to the induced shear (in first order) or flexion (in
second order) of the lens.  Traditional galaxy-galaxy lensing
inversion techniques (Brainerd et al. 1996) yield a relation:
\begin{equation}
P_\phi^{(1)}=\frac{2}{\pi}[1-\langle\gamma\rangle \cos 2\phi_1 \langle
  e^{-1}\rangle]
\label{eq:phi1}
\end{equation}
over the domain $\phi_1=[0,\pi/2]$.  The normalization term, $\langle
e^{-1}\rangle$ is exceedingly noisy, and can be unstable for small
values of ellipticity.  Thus, using the parametric model of Brainerd
et al. (1996):
\begin{equation}
P(e)\propto e \exp{-A\times e}\ ,
\end{equation}
Using this relation, we find $\langle e^{-1}=14.4\rangle$ from the
$\sigma_\gamma$ discussed above.  Note, again, that this is somewhat
larger than the value of $8$ found by Brainerd et al. (1996).  

A virtually identical derivation will yield a relation between the
orientation of the flexion:
\begin{equation}
P_\phi^{(2)}=\frac{1}{\pi}\left[1-\langle A \frac{d\kappa}{dr} \rangle
  \cos\phi_2 \langle  (a_{gal} {\cal F})^{-1}\rangle  \right]\ .
\label{eq:phi2}
\end{equation}
Assuming a similar shape to the shear and Flexion distribution
function, and using the ratio $\sigma_\gamma/\sigma_F=4.1$ found
above, our estimate of the flexion deviation is, $\langle
\left(a_{lens} {\cal F}\right)^{-1}\rangle \simeq 59$, over
$\phi_2=[0,\pi]$.

\section{Proof of Concept: The DLS}
\label{sec:DLS}

As a proof of concept of this technique, we apply it to the Deep Lens
Survey (DLS).  The DLS (Wittman et al. 2002) is an ongoing deep
optical survey of 7$\times$ 4 square degree fields, taken on the NOAO
4m Blanco and Mayall telescopes.  Each field will have an integrated
18ks exposure in the R band, and 12 ks each in the B, V, and z'
bands. For the present test, we use only R band photometry.  While the
DLS is designed for measurements of lensing from Large Scale
Structure, it is also ideally suited for our purposes, as the fields
are selected around otherwise empty regions in the sky, and thus the
primary lensing potential will come from galaxies or large-scale
structure only. 

We looked at all 17 of the DLS subfields from the first three public
data releases, amounting to a total area of approximately 5.4 square
degrees.  We select as potential lensed sources only those galaxies
which have $21.5 < r < 23$, and which have semi-major axes $A >
0.9''$, since objects significantly smaller than that are not
well-resolved into high order shapelets.  This is a far more
conservative cut than on a shallower sample than that used by Brainerd
et al. (1996), however, we wish to re-emphasize that our goal in this
work is to demontrate the detectability of the Flexion signal.  Thus,
taking as potential lenses those galaxies with $18 < r < 21.5$.  In
total, we found 4833 potential pairs (similar to the number found in
the Brainerd et al. 1996 sample) with separations less than 60'' over
17 subfields.

We then decompose all background galaxies into shapelet coefficients
as described above, and estimate the best fit Flexion and shear by
minimization $\chi^2$ as in equation~(\ref{eq:inverse}).  Since seeing
produces a significant reduction in the magnitude of the shear and
Flexion, we use only the orientation angles.  

\begin{figure}[h]
\centerline{\psfig{figure=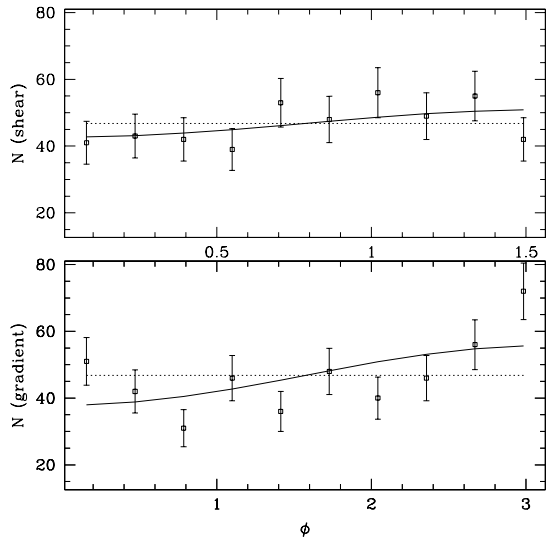,angle=0,height=3in}}
\caption{The distribution of relative orientation angles of the shear
  (top panel) and flexion (bottom panel) with respect to a potential
  lens within a separation of $16''$.  The lines indicate a best fit
  to equations (\ref{eq:phi1}) and (\ref{eq:phi2}), of
  $\gamma=0.006\pm 0.006$ and $d\kappa/dr=0.0027\pm 0.0012$.}
\label{fg:angles}
\end{figure}

The distribution of relative orientation angles for pairs with
separation less than $16''$ and greater than $5''$ is shown in
Fig.~\ref{fg:angles}, along with a best fit curve representing a shear
of $\gamma=0.006\pm 0.006$ and ${\cal F}=0.0027\pm 0.0012 ('')^{-1}$.
We use a KS test in all cases to determine best fit parameters.  
Since the shear and Flexion should be preferentially aligned either
perpendicular (shear), or toward (Flexion) the source, a rotation of
45 degrees for shear and 90 degrees for Flexion (known, hereafter, as
the ``B-field signal'') should produce an entirely random signal.
Applying these rotations to each observed source and computing the
corresponding shear and Flexion produces a means of normalizing the
error bars for our sample, and checking for consistency.

For example, if we divide a given sample into $N_{bin}$ bins of
lens-source separation, then we expect that the B-field terms will
yield a signal with a $\chi^2=N_{bin}-1$, with the errorbars in each
bin being proportional to $\sqrt{rN_i}$, the number of galaxies within
that bin.  Moreover, since the errorbars associated with the B-field
should be the same as for the E-field, we may use this test as a means
of normalising the E-field errorbars. We linearly correct extend the B
error bars (initially assumed to be Poisson noise) such that the
reduced $\chi^2$ is 1. It should be noted that this produces
approximately a 10\% correction in both the Flexion and Shear
errorbars from those based on an analytic estimate of uniformly
sampled orientation angles.

This appears to be assuming that there is no systematic error
contribution to B; however, we are simply applying a useful fiction to
obtain error bars for E that {\it contain} the systematic error
contribution. The method does not explicity find the level of this
systematic error in B, but accounts for the systematic and statistical
errors when quoting an E error.

\begin{figure}[h]
\centerline{\psfig{figure=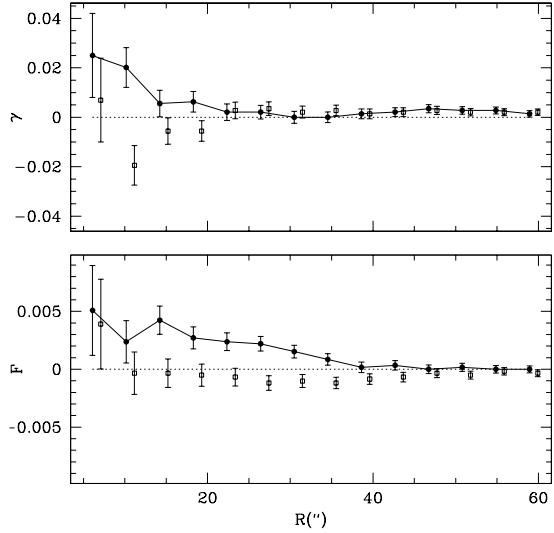,angle=0,height=3in}}
\caption{The cumulative estimate of the shear (top panel) and the
flexion (bottom panel), of the first three releases from the DLS.
Each solid point represents the integrated (from $R_{gal}=5''$)
E-field shear or flexion as described in the text.  Note that the
errorbars on each are strongly correlated.  The figure is plotted such
that positive shear represents source galaxies aligned tangentially to
the lens, and positive values of the flexion actually represent {\it
decreasing} surface density as a function of radius.  The squares
represent the B-field (rotated) signal, and are consistent with no
signal. This cross-signal measurement demonstrates that our observed
signal could not arise randomly.}
\label{fg:bins}
\end{figure}

We then determine an average radial profile for both terms.  We plot
the average cumulative shear (measured within a disk from separations
of $5''$ to r) for both terms in Fig.~\ref{fg:bins}.

The radial profiles represent cumulative estimates of the shear and
flexion, and thus the errorbars are strongly coupled within a given
plot.  At highest significance (around 10'' for the shear, and 20'' for
the flexion), the shear and flexion result in a 1 and 2$\sigma $
detection, respectively.  The shear estimates are also consistent
with estimates from other researchers, notably Brainerd et al. (1996)
who find a shear of $0.0055$ for separations of $R_{lens} < 20''$.

In order to fit the data to physical parameters, we select a simple
isothermal sphere model for the lens, and assume that all lenses are
drawn from the same population.  We then expect the relation
(Bartelmann \& Schneider 2001):
\begin{equation}
|\gamma| =
 \frac{2\pi}{R_{lens}}\frac{v_c^2}{c^2}\frac{D_{ls}}{D_s}\ ,
\end{equation}
and
\begin{equation}
\left|\frac{d\kappa}{dr}\right| =
 \frac{2\pi}{R^2_{lens}}\frac{v_c^2}{c^2}\frac{D_{ls}}{D_s}\ ,
\end{equation}
where $D_{ls}$ is the angular diameter distance from source to the
lens.  We set this ratio to be $0.5$ for our discussion since lensing
signals are typically maximized around this ratio (e.g. Bartelmann \&
Schneider 2003), and will normalize the results to this value.  Note
that an error in this ratio is systematic, as it effects both our
model from the flexion and from the shear identically.

\begin{figure}[h]
\centerline{\psfig{figure=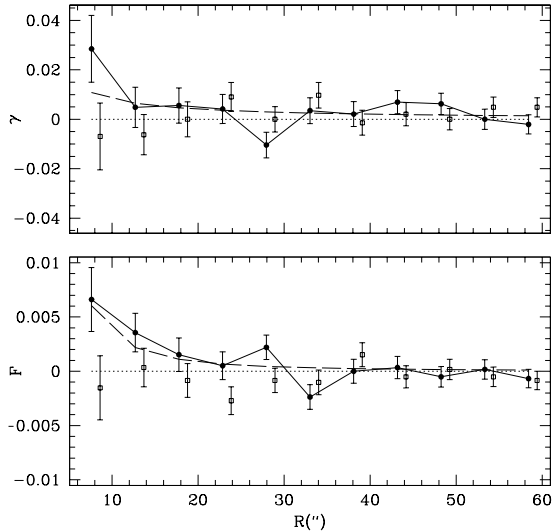,angle=0,height=3in}}
\caption{As in the previous figure, we plot the estimates of the shear
and flexion of galaxies from the DLS.  However, here, we plot the
estimate of the differential shear from annuli of 5''.  We then fit
the curves to an isothermal sphere, as described in the text.  Using
shear, we find $v_c=109$ km/s,, while using Flexion, we find
$v_c=221$ km/s.  Again, the squares show the results of rotating
the sources by 45 degrees for shear and 90 degrees for Flexion.}
\label{fg:fit}
\end{figure}

We fit the isothermal sphere model in Fig.~\ref{fg:fit}.  Note that we
may fit a velocity for either the shear or flexion estimate.  For the
shear, we find a fit of $(v_c=107^{+24}_{-32} km/s)
(D_s/D_{ls}/0.5)^{0.5}$, and for the flexion
$v_c=(209^{+12}_{-13} km/s)(D_s/D_{ls}/0.5)^{0.5}$.  These two
fits may be combined to provide a best fit of $(201\pm 11 km/s)
(D_s/D_{ls}/0.5)^{0.5}$.  While here we are quoting only the random
error, there may also be a systematic error based on errors in the
assumed distribution function of ellipticities and flexions.
The shear estimate produces a significantly lower velocity than that
found by Brainerd et al. (1996; 220 km/s), though the two estimates
straddle the lower velocity estimate of 135 km/s found by Hoekstra et
al. (2004) in the Red Cluster Survey.  Nevertheless, it should be
noted that the shear and Flexion estimate within our own sample are
not statistically consistent.  There are several possibilities.

First, an isothermal sphere may not be the best model.  Consider a
circularly symmetric power-law density field, 
\begin{equation}
\kappa(r)=A r^{-\eta}\ ,
\end{equation}
where $\eta=1$ for an isothermal sphere.  The Flexion for such a lens
will be:
\begin{equation}
|{\cal F}|=A\eta r^{-\eta-1}
\end{equation}
Likewise, symmetry dictates that the shear is:
\begin{equation}
|\gamma|=\overline{\kappa}(r)-\kappa(r)=\frac{\eta}{2-\eta}\kappa(r)
\end{equation}
Thus:
\begin{equation}
\left|\frac{{\cal F}}{\gamma}\right|=\frac{2-\eta}{r}
\end{equation}
Thus, Flexion which indicates a very high velocity dispersion compared
to the shear may, in fact, be indicative of an $\eta < 1$.  In other
words, such a discrepancy may be caused by a density distribution
dropping off more slowly than isothermal.

Secondly, and more likely, recall that though the statistical
significance of the Flexion signal can be determined internally from
the data, the normalization of the signal must be estimated from the
intrinsic distribution, in the case of a KS test.  However, if we've
overestimated the intrinsic variance of the Flexion then we will
overestimate the Flexion as well.  Consider that for fixed
distribution of orientation angles, the normalizations of the curves
may be related as:
\begin{equation}
\frac{v^2_{c,Flexion}}{v^2_{c,shear}}\propto
\frac{\sigma_F}{\sigma_\gamma}
\end{equation}
Thus, the two estimates may be reconciled if the ratio of
$\sigma_F/\sigma_\gamma$ is reduced by a factor of $(107/221)^2=0.23$.
If this is the sole source of the discrepancy, then the Flexion may
dominate the signal over shear on larger scales than originally suggested by
equation~\ref{eq:intrinsic}, yielding the new relation:
\begin{equation}
R_{lens,equality} \simeq 17.8 a_{gal}
\end{equation}
as the crossover scale between Flexion dominance and shear dominance.

\section{Future Prospects}

We believe that we have convincingly laid out and demonstrated the
feasibility of interpreting new information from weak lensing fields
by extending that analysis to second order.  However, the present work
leads to a number of additional areas of investigation both
theoretical and observational.

First, the present work does not incorporate inversion of the Point
Spread Function.  It is therefore clear that we are presently unable to
extract a significant signal from galaxies which are on
order the same size as the PSF.  As a result, we are forced to remove
many potentially lensed galaxies from our sample despite the fact that
information could be potentially extracted from them.  We will address
this issue in future work.

Secondly, the present work has actually thrown away some information
with regards to measurements of the shear derivatives.  Since we have
four derivatives, and reduce that data to a 2-vector (${\bf {\cal
F}}$), we have, in essence, thrown away information.  In principle,
those derivatives could yield, in addition to $d\kappa/dr$, a
measurement of the variations of, say, the shear as a function of
radius.  Since for an isothermal sphere, these numbers are identical
(up to a minus sign), this could potentially be an important test for
isothermality.

As a complimentary effort, we will apply this technique to space-based
data.  Of particular use are the Medium Deep Survey (Ratnatunga,
Griffiths, \& Ostrander 1999) and GOODS (Giavalisco et
al. 2004), which will provide the opportunity to measure more
precisely the intrinsic distribution of flexions in background
galaxies.  Moreover, since the PSF is much smaller, we will be able to
directly measure the component of the shear and flexion parallel to
the lens-source displacement.

In summary, we have detected the second-order lensing effect,
``flexion''. This effect will be of great value on its own and in
conjunction with first-order lensing, in order to measure the
properties of galaxy halos; it affords direct, local information on
the gradient of the halo density.

\acknowledgements
  
DMG acknowledges support from NSF grant AST-0205080, as well as David
Wittman for a number of useful correspondences, and Fiona Hoyle and
Tereasa Brainerd for helpful conversations.  DJB is supported by a
PPARC Postdoctoral Fellowship.

\end{document}